# Attosecond dynamics through a Fano resonance: Monitoring the birth of a photoelectron

**Authors:** V. Gruson[1]†, L. Barreau[1]†, Á. Jiménez-Galan[2], F. Risoud[4], J. Caillat[4], A. Maquet[4], B. Carré[1], F. Lepetit[1], J-F. Hergott[1], T. Ruchon[1], L. Argenti[2]‡, R. Taïeb[4], F. Martín[2,3,5]*, P. Salières[1]*

**Affiliations:**

[1] LIDYL, CEA, CNRS, Université Paris-Saclay, CEA Saclay, 91191 Gif-Sur-Yvette, France.

[2] Departamento de Química, Módulo 13, Universidad Autónoma de Madrid, 28049 Madrid, Spain.

[3] Instituto Madrileño de Estudios Avanzados en Nanociencia (IMDEA-Nanociencia), Cantoblanco, 28049 Madrid, Spain.

[4] Sorbonne Université, UPMC Université Paris 6, UMR 7614, Laboratoire de Chimie Physique-Matière et Rayonnement, 11 rue Pierre et Marie Curie 75231 Paris Cedex 05, France and CNRS, UMR 7614, LCPMR, Paris, France.

[5] Condensed Matter Physics Center (IFIMAC), Universidad Autónoma de Madrid, 28049 Madrid, Spain.

*Correspondence to: fernando.martin@uam.es, pascal.salieres@cea.fr

†These authors contributed equally to this work

‡Current address: Dept. of Physics & CREOL, UCF, Orlando, FL 32816, USA

**Abstract**: Amplitude and phase of wavepackets encode the dynamics of quantum systems. However, the rapidity of electron dynamics on the attosecond timescale has precluded their complete measurement in the time domain. Here, we demonstrate that spectrally-resolved electron interferometry reveals the amplitude and phase of a photoelectron wavepacket created through a Fano autoionizing resonance in helium. Replicas obtained by two-photon transitions interfere with reference wavepackets formed through smooth continua, allowing the full temporal reconstruction, purely from experimental data, of the resonant wavepacket released in the continuum. This in turn resolves the buildup of the autoionizing resonance on attosecond timescale. Our results, in excellent agreement with *ab initio* time-dependent calculations, raise prospects for both detailed investigations of ultrafast photoemission dynamics governed by electron correlation, as well as coherent control over structured electron wave-packets.

**One Sentence Summary:** By monitoring the decay of an excited atom in real time, we reconstruct how photoelectron wavepackets are born and morph into asymmetric Fano profiles.

**Main Text:**

Tracking electronic dynamics on the attosecond (*as*) timescale and Ångström (Å) lengthscale is a key to understanding and controlling the quantum mechanical underpinnings of physical and chemical transformations [1]. One of the most fundamental electronic processes in this context is photoelectron emission, the dynamics of which are fully encoded in the released electron wavepacket (EWP) and the final ionic state. The development of broadband coherent sources of attosecond pulses has opened the possibility of investigating these dynamics with attosecond resolution. On such a short timescale, few techniques [2-5] are able to provide access to both spectral amplitude and phase. The spectral derivative of the phase, the group delay, is a practical quantity for describing general wavepacket properties reflecting the ionization dynamics. Recently, photoemission delays have been measured in a variety of systems: rare gas atoms [6-8] molecules [9] and solids [10]. In the gas phase, these attosecond delays give insight into the scattering of the electron in the ionic potential, and in the solid state, into the transport dynamics towards the surface. However, the physical significance of group delays is restricted to fairly unstructured wave-packets.

The necessity to go beyond simple delays arises for more complex ionization dynamics when the broadband excitation encompasses continuum structures associated with, e.g., autoionizing states, shape resonances, and Cooper minima [11-13]. These structures induce strong spectral variations of the amplitude and phase of the EWP corresponding to different timescales, e.g., ranging from the attosecond to the femtosecond domains. In general, the long-term evolution of the EWP amplitude, e.g., the lifetime of Fano autoionizing resonances [14], can be characterized directly in the time-domain [15], or in the spectral domain using conventional spectroscopic techniques [16]. However, the EWP phase is required in order to reconstruct the full ionization dynamics. In particular, the short-term response associated with broadband excitation remains unexplored [17]. It is mainly determined by the spectral phase variation over the resonance bandwidth, which has so far not been measured. An additional difficulty is that the characterization techniques usually involve strong infrared probe fields that: i) strongly perturb the resonant structures [18-20] so that the field-free intrinsic dynamics cannot be accessed, and ii) require elaborate theoretical input for decoding the electron spectrograms [21].

Here, we extend attosecond photoionization spectroscopy to the full reconstruction of the time-dependent EWPs produced by coherent broadband excitation through resonant structures. To this end, we develop a perturbative interferometric scheme enabling the direct measurement of the spectral amplitude and phase of the unperturbed resonant EWP. Interferences between the latter and a reference non-resonant EWP are achieved through two-photon replicas obtained by photoionizing the target with an XUV harmonic comb combined with the mid-infrared (MIR) fundamental field. This spectrally-resolved technique is easy to implement and offers straightforward access to the EWP characteristics without complex analysis or theoretical input. We apply it to the investigation of the doubly-excited 2s2p autoionizing resonance of helium, for which *ab initio* time-dependent calculations can be performed [22,23] providing a benchmark for our experimental study. Autoionization occurs when a system is excited in structured spectral regions where resonant states are embedded into a continuum. The system can then either directly ionize or transiently remain in the resonant bound state before ionizing. Coupling between the resonant state and continuum states of the same energy through configuration interaction leads to the well-known Fano spectral lineshapes [14]. Particularly interesting is the autoionization decay from doubly excited states [16] that is a direct consequence of the electron-electron repulsion. Using our spectrally resolved technique, we directly access the complete ionization dynamics including interferences at birth time, and monitor the resonance buildup on a sub-fs timescale, a long-awaited endeavor of attosecond science [17,24].

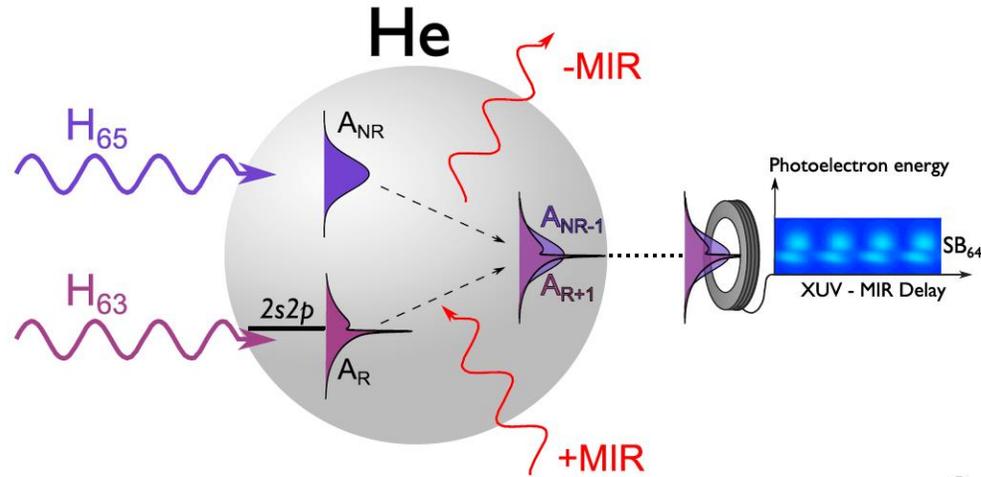

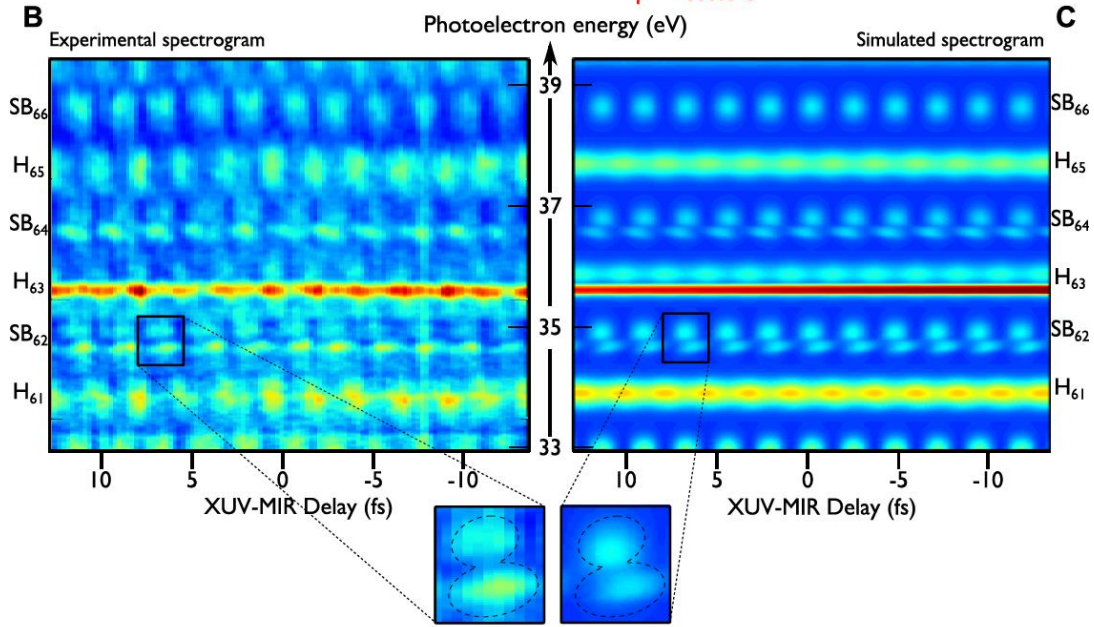

**Fig. 1. Principle and resulting trace of spectrally-resolved attosecond electron interferometry for the complete characterization of resonant EWPs.** (**A**) Principle of the electron interferometry technique: resonant $A_R$ and reference non-resonant $A_{NR}$ EWPs are produced by successive coherent harmonics. Replicas of these EWPs are created at the same final energy by 2-photon transitions induced by a weak fundamental MIR field, where the atom absorbs a MIR photon leading to the $A_{R+1}$ EWP, or emits a MIR photon, leading to the $A_{NR-1}$ EWP. The spectrally-resolved interferences measured in a time-of-flight electron spectrometer as a function of the XUV-MIR delay $\tau$, controlled with interferometric accuracy, provide access to the spectral phase of the resonant $A_R$ EWP, considering the non-resonant $A_{NR}$ as a reference. (**B**)

Experimental and **(C)** theoretical spectrograms in the 33 to 39 eV region for a 1295-nm OPA wavelength [25]. $H_{63}$ overlaps the 2s2p resonance of helium located 60.15 eV above the ground state ($E_R$=35.55 eV). Single-photon ionization by the odd harmonic orders results in main lines spaced by twice the MIR photon energy, $2\hbar\omega_0$ =1.92 eV. In-between appear sidebands corresponding to 2-photon ionization. The oscillations of the two sidebands on both sides of the resonant $H_{63}$, i.e. $SB_{62}$ and $SB_{64}$, encode the spectral phase of the resonant EWP. A blow-up of one $SB_{62}$ beating shows the structured shape of this resonant EWP and the dephasing of the oscillations of the different spectral components.

The concept of the method is shown in Fig. 1A. We photoionize helium with a comb of mutually coherent odd harmonics derived from an optical parametric amplifier (OPA) MIR source. The harmonic of order 63 ($H_{63}$) is driven into the 2s2p resonance, at 60.15 eV from the ground state, by tuning the OPA central wavelength $\lambda_{OPA}$ to 1295 nm. As the harmonic width (400 meV) is much larger than the resonance width ($\Gamma$=37 meV), a broad resonant EWP with complex spectral amplitude $A_R(E)$ is produced. Simultaneously, non-resonant EWPs are created by the neighboring harmonics $H_{61}$ and $H_{65}$ in smooth regions of the continuum: each of these can serve as a reference, denoted $A_{NR}(E)$, to probe the resonant EWP. In order to induce interference, we create replicas that spectrally overlap with each other, by means of two-photon transitions. A weak fraction of the fundamental MIR pulse, of angular frequency $\omega_0=2\pi c/\lambda_{OPA}$, is superimposed on the harmonic comb with a delay $\tau$. Its intensity (~2 $10^{11}$ W/cm$^2$) is sufficiently high to induce perturbative 2-photon XUV-MIR transitions but is low enough to avoid transitions involving more than 1 MIR photon (e.g. depletion of the doubly excited state by multiphoton ionization [15], or distortion of the resonance lineshape [19]). Most importantly, the MIR spectral width (26 meV) is smaller than both the harmonic and resonance widths, ensuring that each EWP produced in the two-photon process is a faithful, spectrally shifted, replica of the unperturbed EWP produced in the XUV one-photon process. Due to the frequency relation between the odd-harmonic XUV comb and the fundamental MIR laser, the resonant EWP up-shifted by absorption of a MIR photon, $A_{R+1}(\tau, E + \hbar\omega_0) \propto A_R(E) \exp i\omega_0\tau$, and the reference EWP down-shifted by stimulated emission of a MIR photon, $A_{NR-1}(\tau, E + \hbar\omega_0) \propto A_{NR}(E + 2\hbar\omega_0) \exp[-i\omega_0\tau]$, coherently add up in the single sideband ($SB_{64}$) that lies in between the lines associated with $H_{63}$ and $H_{65}$. Similarly, the resonant EWP down-shifted by emission of a MIR photon interferes in sideband $SB_{62}$ with the EWP up-shifted by absorption of a MIR photon from $H_{61}$. We designate $E$ the photoelectron energy in the resonant EWP, and $\bar{E} = E \pm \hbar\omega_0$, the photoelectron energy of the resonant EWP replicas in $SB_{64}$ and $SB_{62}$, respectively.

The spectrum of these SBs is thus modulated by the interference between the resonant and non resonant replicas, depending on the XUV-MIR delay $\tau$ [25]. For $SB_{64}$, it gives (eq. 1):

$$S_{64}(\tau, \bar{E}) = |A_{R+1}(\tau, \bar{E}) + A_{NR-1}(\tau, \bar{E})|^2 = |A_{R+1}(\bar{E})|^2 + |A_{NR-1}(\bar{E})|^2 \\ +2|A_{R+1}(\bar{E})||A_{NR-1}(\bar{E})| \times \cos\{2\omega_0\tau + \Delta\varphi_{64}(\bar{E}) + \Delta\eta_{scat}(\bar{E})\}, \quad (1)$$

where the two contributions to the replicas' relative phase are: i) $2\omega_0\tau$, the phase introduced by the absorption/emission of the MIR photon, and ii) the relative phase between the initial one-photon EWPs. The latter is split into: $\Delta\varphi_{64}(\bar{E}) = \varphi_{65}(\bar{E} + \hbar\omega_0) - \varphi_{63}(\bar{E} - \hbar\omega_0)$, the phase difference between the two ionizing harmonics, and $\Delta\eta_{scat}(\bar{E}) = \eta_{scat}(\bar{E} + \hbar\omega_0) - \eta_{scat}(\bar{E} - \hbar\omega_0)$, the difference between the non-resonant and resonant scattering phases of the two intermediate states. In our conditions, the variation over the SB width of both $\Delta\varphi_{64}(\bar{E})$ and $\eta_{scat}(\bar{E}+\hbar\omega_0)$ is negligible in comparison with that of the resonant scattering phase $\eta_{scat}(\bar{E}-\hbar\omega_0)$ [25]. The latter contains information on the scattering of the photoelectron by the remaining core, including strongly correlated scattering by the other electron close to the resonance. This is the observable addressed by our study.

Using a high-resolution (~1.9%) 2m-long magnetic-bottle spectrometer, we have access to the photoelectron spectrogram –electron yield as a function of energy $E$ and delay $\tau$– spectrally resolved within the harmonics and sidebands widths, as presented in Fig. 1B. Due to its large bandwidth, H$_{63}$ produces a photoelectron spectrum exhibiting a double structure with a resonant peak and a smoother peak. This shape is replicated on each of the closest resonant SBs (SB$_{62}$ and SB$_{64}$). Strikingly the components of the double structure oscillate with different phases when $\tau$ is varied, in both SB$_{62}$ and SB$_{64}$.

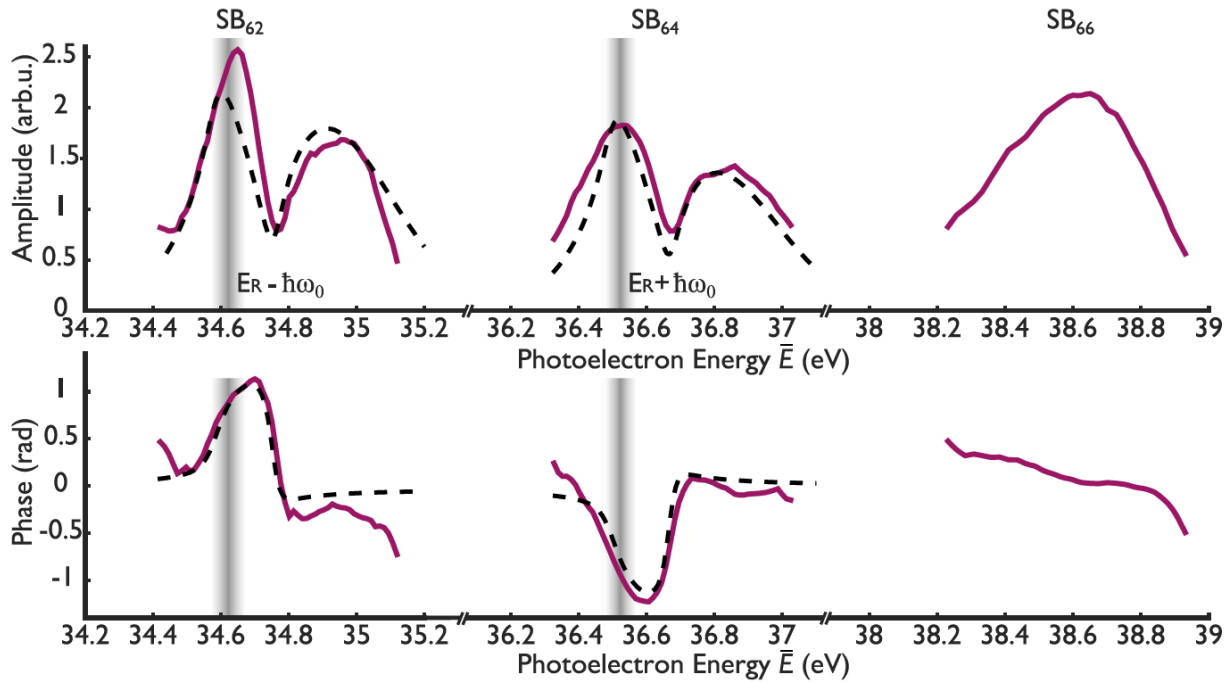

**Fig. 2. Resonant EWP in the spectral domain.** Spectral amplitude (upper panels) and phase (lower panels) of the $2\omega_0$ component of SB$_{62}$ (left), SB$_{64}$ (middle), and SB$_{66}$ (right) from the spectrograms in Figs. 1B-C. The phase origin is set to 0 by removing the linear variation due to the ionizing harmonic radiation (atto-chirp) [26]. The experimental data (purple curves) are compared to the simulations (dashed black lines), showing very good agreement. The resonance position shifted by one MIR photon is indicated in grey. The measured spectral amplitudes and phases of the resonant SB$_{62}$ and SB$_{64}$ are easily related to the amplitude $|A_R(E)|$ and phase $\eta_{scat}(E)$ of the resonant one-photon EWP (see eq. 1). The main limitation comes from the current spectrometer resolution (in our conditions, a relative resolution of ~1.9% resulting in ~190-meV width at 10 eV) that broadens the resonant peak and its phase variations. The non-resonant SB$_{66}$ exhibits a Gaussian amplitude, which mostly reflects the ionizing XUV spectral profile, and a smooth close-to-linear phase. This provides a temporal reference for the ionization dynamics.

These phase variations are further evidenced by a spectrally-resolved analysis: for each sampled energy within the SB width, we perform a Fourier transform of $S_{63\pm1}(\tau, E \pm \hbar\omega_0)$ with respect to $\tau$ to extract the amplitude and phase of the component oscillating at $2\omega_0$ (see eq. 1 and Fig. 2). The $SB_{62}$ phase shows a strong increase of ~1 rad within the resonant peak, followed by a sudden drop at the amplitude minimum ($\bar{E}$~34.75 eV), and a rather flat behavior under the smooth peak. The $SB_{64}$ phase has a very similar shape and magnitude but with an opposite sign due to opposite configuration of the resonant and reference EWPs in the interferometer. This correspondence confirms the direct imprint of the intermediate resonance on the neighboring sidebands.

The $2\omega_0$ component of the resonant sidebands thus provides a good measure of the $|A_R(E)|\exp(i\eta_{scat}(E))$ EWP that would result from 1-photon Fourier-limited excitation. This allows a detailed study of the temporal characteristics of resonant photoemission, in particular of the electron flux into the continuum, through the direct reconstruction of this EWP in the time domain (eq. 2):

$$\tilde{A}_R(t) = \frac{1}{2\pi}\int_{-\infty}^{+\infty}|A_R(E)|e^{i\eta_{scat}(E)}e^{\frac{-iEt}{\hbar}}dE. \qquad (2)$$

The temporal profile obtained from $SB_{64}$ is shown in Fig. 3A. It presents a strong peak at the origin – given by the maximum of the Fourier transform of the non-resonant $SB_{66}$ [25]– followed by a deep minimum around 4 fs and then a revival with a decay within ~10 fs. The presence of a fast phase jump (~2 rad within ~2 fs) at the position of the minimum indicates that it results from a destructive interference between two wave-packet components, the origin of which will be detailed further.

To benchmark the measured data, we theoretically investigated the multicolor XUV+MIR ionization of He in the vicinity of the 2s2p resonance. Fully correlated *ab initio* time dependent calculations [22] were used to validate an analytical model of the two-photon transitions accounting for the actual pulses' bandwidths [23]. The simulated photoelectron spectrogram taking into account the spectrometer resolution remarkably reproduces the structured shape of the resonant SBs, as well as the dephasing between their two components (Fig. 1B). The analysis of the $2\omega_0$ oscillations of $SB_{62}$ and $SB_{64}$ gives spectral phase variations in excellent agreement with the experimental data (Fig. 2). The temporal profile $\tilde{A}_R(t)$ obtained by Fourier transform (Fig. 3) is also well reproduced, with a smaller revival but a similar decay time of ~10 fs. This reduced effective lifetime is a direct consequence of the finite spectrometer resolution. When the latter is assumed infinite, the time-profile has the same behavior at short times but a longer decay, corresponding to the 17 fs lifetime of the resonance. Analytical calculations show that, in our conditions, the reconstructed EWP does mirror the one-photon resonant EWP [25]. In summary, this confirms that, except for a faster decay of the long-term tail due to our current electron spectrometer resolution, the essential physics of the early time frame of EWP creation is directly accessed from purely experimental data.

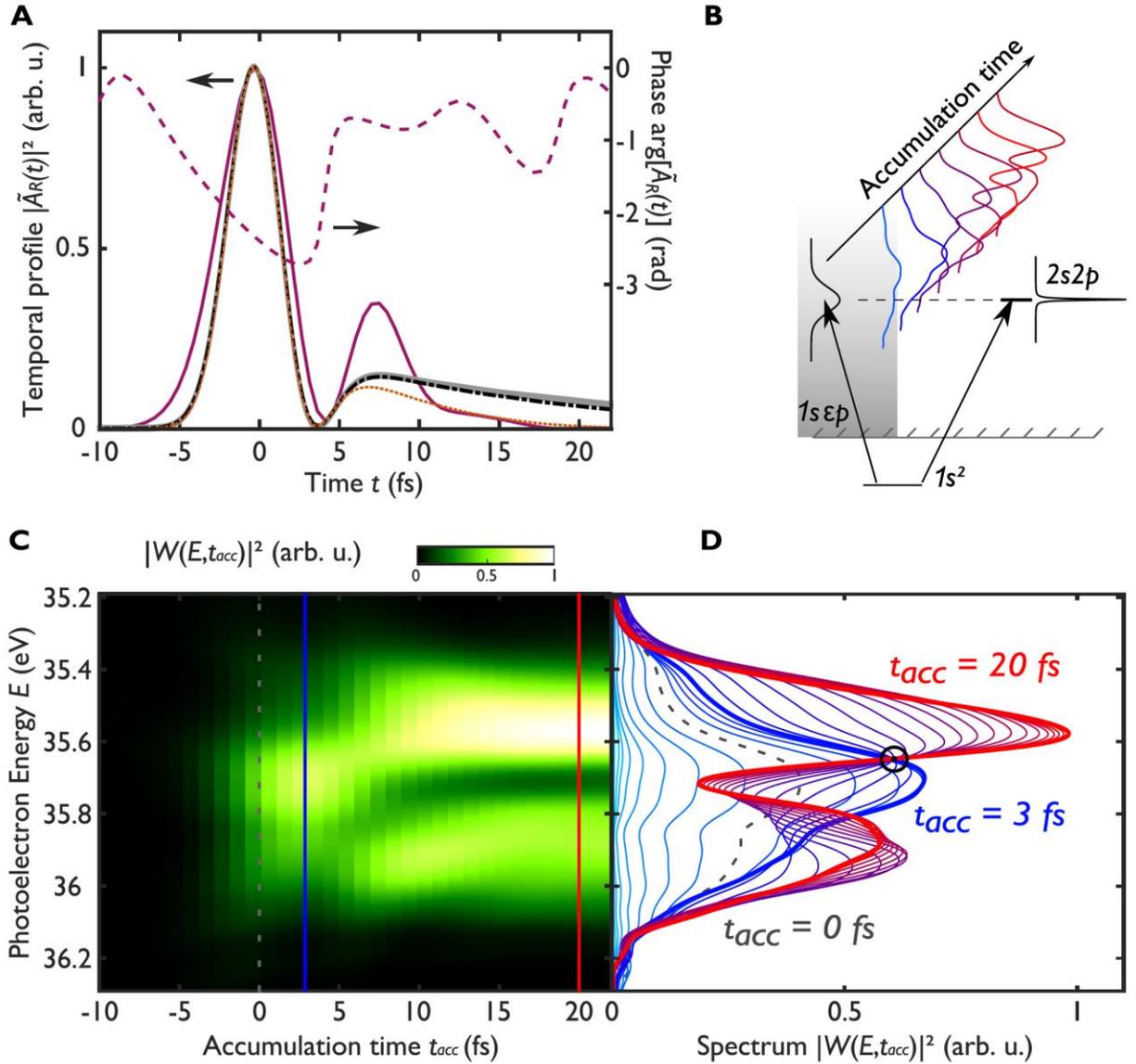

**Fig. 3. Resonant EWP in the time domain and time-resolved reconstruction of the resonance build-up.** **(A)** Temporal profile of the resonant EWP obtained by Fourier transform of the $SB_{64}$ data: i) from the experimental spectrogram (solid purple), and corresponding temporal phase (dashed purple), ii) from the simulated spectrogram taking into account (dotted orange) or not (dot-dashed black) the finite spectrometer resolution. The latter fully coincides with the one-photon resonant EWP profile from a direct analytical calculation (solid grey) [25], demonstrating the validity of our interferometric technique. **(B)** Illustration of the formation dynamics of the resonant spectrum resulting from interference between the two paths in the Fano autoionization model. **(C)** Reconstruction of the time-resolved buildup of the resonant spectrum using the time-energy analysis introduced in eq. 5. The photoelectron spectrum is plotted as a function of the upper temporal limit (accumulation time $t_{acc}$) used for the inverse Fourier transform. **(D)** Lineouts of (C) every 1 fs. This figure evidences first the growth of the direct path until a maximum is reached at ~3 fs (blue), and then the increasing spectral interference with the resonant path that finally results in the Fano lineshape (red). At 35.6 eV, an isosbestic-like point

is crossed by all curves from 3 fs onwards (black circle), evidencing a position in the final lineshape where only the direct path contributes.

To further highlight the insight provided by this experimental technique, we undertook an in-depth analysis of the measured EWP characteristics in terms of Fano's formalism for autoionization [14]. Resonant ionization can be described as the interference between two distinct paths: the direct transition to the continuum and the resonant transition through the doubly-excited state that eventually decays in the continuum through configuration interaction within the resonance lifetime (Fig. 3B). The normalized total transition amplitude can then be written as the coherent sum of two contributions, a constant background term and a Breit-Wigner amplitude for the resonance (eq. 3):

$$R(E) = \frac{\epsilon+q}{\epsilon+i} = 1 + \frac{q-i}{\epsilon+i}, \tag{3}$$

where $\epsilon = 2(E - E_R)/\Gamma$ is the reduced energy detuning from the resonance at energy $E_R$, in units of its half width $\Gamma/2$. The Fano parameter $q$ (-2.77 for the He(2s2p) resonance [16]) measures the relative weight of the two paths. Their interference leads to the well-known asymmetric Fano lineshape $|R(E)|^2$ and to the resonant scattering phase: $\eta_{scat}(E) = \arg R(E) = \mathrm{atan}(\epsilon) + \pi/2 - \pi\Theta(\epsilon + q)$ where $\Theta$ is the Heaviside function. This phase is experimentally accessed here (Fig. 2).

The spectral amplitude of an EWP created by Gaussian harmonic excitation $H(E)$ is given by $R(E)H(E)$. Its temporal counterpart is $\tilde{A}_R(t) = [\tilde{R} * \tilde{H}](t)$, where $\tilde{R}(t)$ and $\tilde{H}(t)$ are Fourier transforms of the spectral amplitudes, in particular (eq. 4) [24]:

$$\tilde{R}(t) = \delta(t) - i\frac{\Gamma}{2\hbar}(q-i)e^{-\left(\frac{iE_R}{\hbar}+\frac{\Gamma}{2\hbar}\right)t}\Theta(t) \tag{4}$$

The temporal profile $\tilde{A}_R(t)$ thus decomposes into a Gaussian non-resonant term and a resonant contribution, like our experimental data (Fig. 3A). The destructive temporal interference between the two terms leads to the amplitude minimum and phase jump identified around $t = 4$ fs.

To illustrate how the interference between the two paths governs the formation of the resonance lineshape, Wickenhauser et al. [17] introduced a time-frequency analysis based on the limited inverse Fourier transform (eq. 5):

$$W(E, t_{acc}) = \int_{-\infty}^{t_{acc}} \tilde{A}_R(t) e^{iEt/\hbar} dt \tag{5}$$

showing how the spectrum builds up until accumulation time $t_{acc}$. The result of this transform applied to the experimental EWP shown in Fig. 3A is presented in Figs. 3C-D. The chronology of the resonance formation can be nicely interpreted within Fano's formalism. In a first stage until ~3 fs, a close-to-Gaussian spectrum reflecting the ionizing harmonic spectral shape emerges: the direct path to the continuum dominates. Then the resonant path starts contributing as the populated doubly excited state decays in the continuum: interferences coherently build up until ~20 fs, consistent with the temporal profile in Fig. 3A, to eventually converge towards the asymmetric measured spectrum. The resonance growth can thus be decomposed in two nearly consecutive steps governed by fairly different time scales.

The build-up of the resonant profile reveals the presence of a notable point around $E = 35.6$ eV where, as soon as the direct ionization is completed, the spectrum barely changes with $t_{acc}$ any

longer. This can be explained by splitting the $|R(E)|^2$ spectrum from eq. 3 into three terms [27] (eq. 6):

$$|R(E)|^2 = 1 + \frac{q^2+1}{\epsilon^2+1} + 2\frac{q\epsilon-1}{\epsilon^2+1} \qquad (6)$$

At this isosbestic-like point, i.e. for $\epsilon = (1/q - q)/2$, the bound (second) and coupling (third) contributions ultimately cancel each other, leaving only the direct continuum contribution (first term). This point thus gives a useful landmark in the resonant lineshape, e.g., for cross section calibration or reference purposes.

Spectrally-resolved electron interferometry thus provides insight into the ultrafast strongly correlated multielectron dynamics underlying autoionization decay. Given the generality and wide applicability of the Fano formalism (see, e.g. [27]), we anticipate that our approach combined with progress in attosecond pulse production and particle detection (e.g., access to photoelectron angular distributions) will open prospects for studies of complex photoemission dynamics close to resonances and, more generally, structured EWP dynamics in a variety of systems, from molecules [28-30] and nanostructures [27] to surfaces [10]. Furthermore, the well-defined amplitude and phase distortions induced by the resonance offer a means for shaping the broadband EWP, bringing opportunities for coherent control in the attosecond regime.

**Acknowledgements:** We thank S. Weber for crucial contributions to the PLFA attosecond beamline, D. Cubaynes, M. Meyer, F. Penent, J. Palaudoux, for setup and test of the electron spectrometer, and O. Smirnova, for fruitful discussions. Supported by ITN-MEDEA-641789, ANR-15-CE30-0001-01-CIMBAAD, ANR11-EQPX0005-ATTOLAB, the European Research Council Advanced Grant XCHEM no. 290853, the European COST Action XLIC CM1204, and the MINECO Project no. FIS2013-42002-R. We acknowledge allocation of computer time from CCC-UAM and Mare Nostrum BSC.